Review Article

Mustapha El Moussaoui*

# Architectural practice process and artificial intelligence – an evolving practice

**Abstract:** In an era of exponential technological advancement, artificial intelligence (AI) has emerged as a transformative force in architecture, reshaping traditional design and construction practices. This article explores the multifaceted roles of AI in the architectural process, emphasizing its potential to enhance creativity and efficiency while addressing its limitations in capturing multisensory and experiential dimensions of space. Historically, architectural innovation has paralleled technological progress, from basic tools to advanced computer-aided design systems. However, the integration of AI presents unique challenges, requiring architects to critically evaluate its role in design. A narrative review methodology was adopted, focusing on academic sources selected for their relevance, recency, and credibility. The findings reveal that AI is increasingly integrated across various stages of the architectural process, from early conceptualization and site analysis to generative design and construction detailing. AI tools excel at automating repetitive tasks and generating innovative design solutions, freeing architects to focus on creativity and problem-solving. Additionally, AI's (text-to-image) visual representation strength challenges the ocular-centric approaches in architecture, which should push future architects to address the holistic sensory and experiential qualities of space or the critical thinking inherent to architectural design. While AI offers transformative potential, architects must view it as a collaborative partner rather than a passive tool. Integrating AI into design paradigms that prioritize sensory richness, critical thought, and sustainability is essential for creating spaces that resonate on a deeper, more human level. By adopting a balanced relationship between human creativity and AI's computational power, the architectural profession can benefit from this technological opportunity without compromising its core values.

# 1 Introduction

People have often been worried about the arrival of new technologies, especially those that have the potential to change the way things have always been done. When computer-aided design (CAD) tools like AutoCAD were first introduced in the 1980s, architects had similar worries [1]. Many experts were worried that these tools would take the place of architects and designers, making their jobs and skills less important. However, as the industry adapted, it became clear that CAD tools did not replace architects but instead enhanced their capabilities, allowing them to work more efficiently and explore new design possibilities [2]. Despite this, there is a need for a new generation of CAD systems that can respond to issues such as ambiguity, discontinuity, and nonmonotonicity in shape interpretation [3].

The architecture, engineering, and construction industry has been slow to fully integrate computer-based techniques into its business processes, but developments such as object technology, industry standardization initiatives, and the adoption of Internet technologies are driving the need for restructuring [4]. Consequently, large areas and offices were no longer needed for architects to manually draw. Moreover, in terms of project speed, production, and scale, smaller architectural practices that simultaneously invest in CAD systems could rival much larger and more established practices [5]. Large companies drastically shrank or went out of business as a result, drastically altering the industry. Rewinding to the year 2000, building information modelling (BIM) was starting to make a name for itself as the upcoming disruptive technology. Concurrently, pioneers in the field of a new geometrically sophisticated style of architecture included Frank Gehry and Zaha Hadid. Their adoption of newly developed parametric and procedural computer modeling software, like CATIA, Rhino, and Grasshopper, allowed for innovation in building form. Design and construction dynamics were impacted even before the advent of the mobile phone [6]. The purpose of this brief contextualization is to show that technological change does not shield or shield the creative industries from it. In fact, in the past, less advanced technology than the current wave of robotics, algorithms, and AI has the potential to seriously disrupt them.

* **Corresponding author: Mustapha El Moussaoui,** Faculty of Design and Art, Libera Universita di Bolzano, Bolzano, Italy,
e-mail: elmoussaoui.mh@gmail.com

Today, the architectural field is facing another technological paradigm shift as artificial intelligence (AI) continues to develop and infiltrate various aspects of the design and construction process [7]. Similar to the initial response to CAD, there is a mix of excitement and apprehension about the potential impacts of AI on the profession. This article aims to provide an overview of AI's current integration into the architectural process, discuss its potential to support design practices, and emphasize the importance of maintaining a human–AI collaborative approach. It will argue how an architectural domain is changing, and like CAD, AI is not a replacement for human expertise and creativity but rather a tool that should be utilized. However, this paper will also conclude on the risk potential for the whole industry, from education to construction, if we continue approaching the architectural process in the same way we have been doing for the last decade.

## 2 Methodology

This article aims to provide a comprehensive overview of the integration of AI into the architectural design process. By examining the current applications and implications of AI technologies, the article seeks to highlight how AI is already embedded in various stages of the architectural process and to explore the necessity for architects to adapt to these advancements.

To achieve these objectives, a comprehensive narrative review approach was adopted. The research began with an academic database search, including Google Scholar, JSTOR, and IEEE Xplore, using keywords such as "AI in architecture," "AI design tools," "AI in construction," and "architectural innovation." The selection criteria focused on relevance, recency (publications from the last 5 to 10 years), and credibility (peer-reviewed articles, reputable industry reports, and case studies).

The selection of AI tools for this study was guided by the rapid pace of technological development, with new tools emerging continuously. Given this dynamic landscape, it is impractical to cover every available tool. Therefore, the AI tools chosen for analysis were selected based on their prominence in the industry, innovative capabilities, and substantial representation in academic and online articles. Moreover, the tools mostly investigated are those that span from the preliminary design phase into the construction documents, which actually most of the AI tools, are in these representational and data analytic phases. In total, more than 70 AI tools were analyzed during the study.

Key information was extracted from the selected sources, focusing on AI technologies used in architecture, their benefits, challenges, implications, and real-world applications.

The findings were synthesized to offer a structured overview of AI applications, highlighting benefits such as enhanced design capabilities and increased efficiency. Furthermore, given that academic research often lags behind the rapid advancements in the AI industry, a review and analysis of YouTube videos and popular media platforms were conducted. These videos showcased architects utilizing various AI tools, providing insights into the latest technologies and methodologies employed in the field.

## 3 History and the use of new technology in architecture

From the invention of the compass and T-square to the development of CAD software and, more recently, AI applications, the history of architecture is tied to the development of technology. With each new technology, architects have had to change and learn new skills, and they have often had to deal with resistance and skepticism from other architects.

When CAD tools came out in the 1980s [1], it was a big change in the field of architecture. Architects went from using hand-drawn plans and physical models to using digital design processes [5]. The introduction of CAD tools was met with resistance due to concerns about job security and the potential devaluation of human creativity and skill [8,9]. However, studies have shown that the use of CAD tools can significantly improve productivity and job satisfaction [9]. Despite this, many users have been found to underutilize the available resources, suggesting a need for active assistance in the form of spontaneous advice and relevant information [10]. The impact of CAD tools on creative problem-solving in engineering design has been found to be both positive and negative, with the potential for enhanced visualization and communication, but also the risk of circumscribed thinking and premature design fixation [11].

The adoption of CAD tools within architectural practices marked a significant shift from traditional methods,

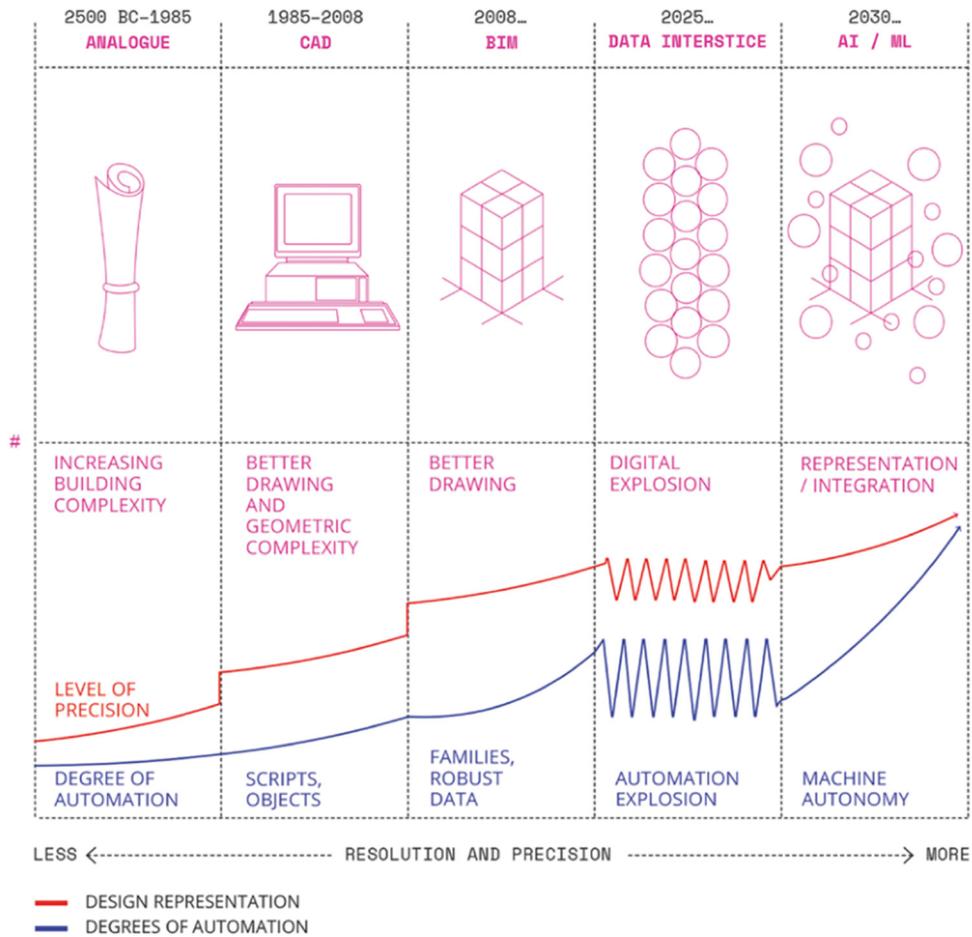

**Figure 1:** Architectural tool advancing through the ages. Source: Bernstein (2022) [85].

significantly broadening the potential for architects to enhance their productivity and innovate in their design approaches. However, this transition also introduced complexities, notably in the perception and manipulation of space. Traditional drafting, with its reliance on fixed scales, offered a tangible sense of scale and dimensions, a feature that became more abstract with the digital interface of CAD systems. This evolution toward a more fluid and dynamic understanding of space reflects a fundamental transformation in architectural cognition and design processes [11].

The historical introduction of CAD necessitated a reevaluation of spatial apprehension, moving away from the constraints of physical media to the limitless possibilities afforded by digital environments. This shift did not simply represent a change but initiated a continuing evolution in architectural design, where the understanding and conceptualization of space are in constant flux, influenced by ongoing advancements in digital technology [10]. The question remains not if this represents a change or a continuum but how architects can navigate and leverage these evolving paradigms to utilize innovation while maintaining a comprehensive grasp of spatial dynamics. The lessons learned from this period of technological adoption can provide insights for the current integration of AI into the field of architecture.

Machine learning technologies are the next technical advancement for architects, coming after BIM. In the initial millennia, our technology primarily relied on analog methods, when architects conveyed design information by markings on paper. In the late 1980s, the manual process was replaced with computerized methods, specifically CAD, to create the designs. By the mid-2000s, there was a notable change in the way information was represented in the construction industry. This change was the widespread availability and adoption of BIM. Currently, professionals are experiencing a situation known as the "digital interstice," where certain aspects of the design and construction process are occasionally converted into digital form using different software and data formats. However, this transformation is happening without a clear structure or organization, which is a common characteristic of the industry. AI, particularly algorithms and skills based on machine learning, are on the verge of being widely available in architecture, while their usage is currently limited [12].

The integration of machine learning tools signals the next frontier, posing both challenges and opportunities. As with the transition to CAD and BIM, the adoption of AI requires a reevaluation of existing processes and a willingness to embrace new methods of design and construction (Figure 1).

As AI technologies advance and increasingly integrate into the design and construction sectors, the architectural profession encounters renewed apprehension regarding the future role of architects. This concern primarily revolves around the potential of AI to substitute for human architects or to diminish the relevance of their skills. Specifically, the skills in question encompass both technical competencies, such as drafting, modeling, and computational design, and aesthetic sensibilities, including the ability to conceptualize and realize visually compelling and spatially harmonious structures. These aesthetic skills are fundamental to the architect's role in creating designs that resonate on a human level, blending form, function, and environmental context into cohesive and meaningful spaces. The apprehension is not unfounded, as AI applications in architecture, such as generative design algorithms, have demonstrated capabilities to produce design solutions based on specific parameters and objectives. However, while AI can assist in optimizing designs for efficiency, sustainability, or cost, the nuanced understanding of aesthetic value, cultural significance, and emotional impact remains a distinctly human domain. Architectural design is as much an art as it is a science, and the creative intuition and judgment architects bring to a project are irreplaceable by current AI technologies [13,14].

Therefore, the challenge lies not in the replacement of architects by AI but in how architects can leverage AI as a tool to augment their skills, including aesthetic judgment. This involves a symbiotic relationship where AI handles more quantitative and data-driven aspects of design, allowing architects to focus more on qualitative and creative aspects, thus enhancing the overall design process rather than diminishing the architect's role [7,15].

However, examining AI presents unique challenges compared to other subjects. The mix of hype, misconceptions, and the inherent complexity of AI creates a barrier that keeps the field of architecture warily at bay from fully embracing this technology. Given its somewhat peripheral relevance to architectural practice, it begs the question: Why should architects invest interest in AI.

In a report by Goldman Sachs titled "The Potentially Large Effects of AI on Economic Growth" underscores the prospective economic impact of AI on areas ranging from office and administration support to cleaning services. The architecture and engineering industry is set to have 37% of the market that is being affected [16] (Figure 2).

In addition to the report, another primary insight into this question is the concept of "projection," a key aspect of AI. At its core, projection involves transferring a specific element from one area to another, essentially calculating or constructing these transfers either mathematically or geometrically. The development of perspective in architecture serves as a timeless marker of projection's significance, enabling the creation of three-dimensional (3D) views from basic geometric principles based on plans

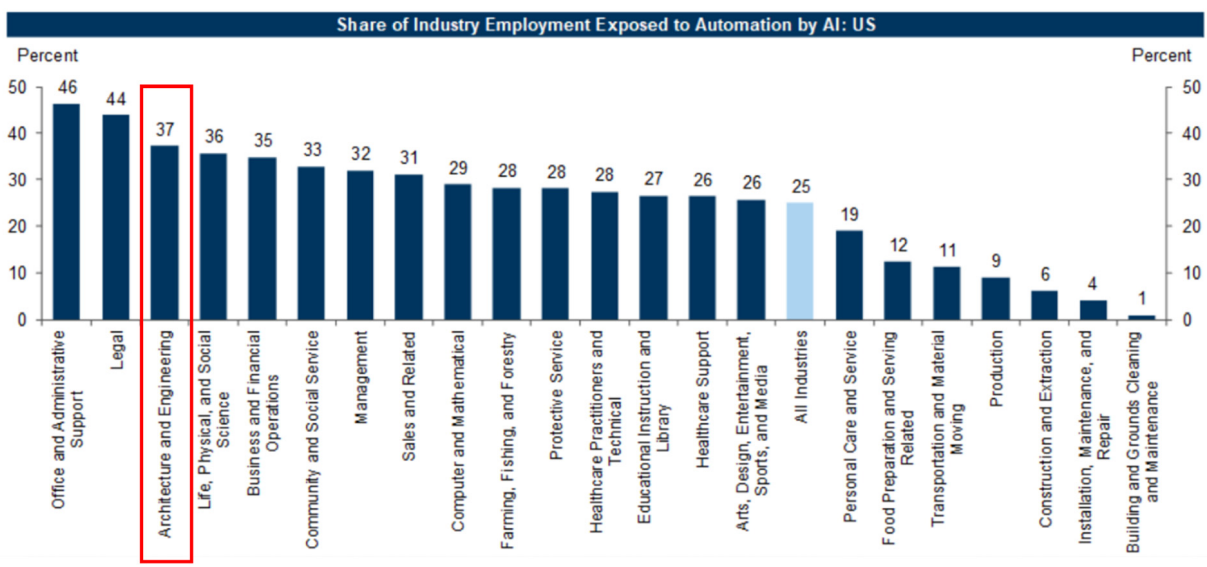

*Source:*

**Figure 2:** Goldman Sachs report on the share of industry employment exposed to automation by AI. Source: Goldman Sachs (2023).

and elevations. Historical figures like Gaspard Monge have significantly advanced descriptive geometry, enhancing the toolkit available for managing spatial information across various fields.

However, the evolution of projective methods has significantly benefited architecture and numerous related disciplines. The integration of spatial transformations into architectural methods highlights the ongoing impact of computer technology in modern times. This underscores the critical role of projection in the field, a sentiment shared by both theorists and practitioners alike.

This importance is further underscored by the rapid advancements in projective geometric techniques driven by AI. Projects like NeRF exemplify this advancement, demonstrating the ability to create full 3D models from limited two-dimensional (2D) views, revealing unseen geometries. This leap from partial to complete representations signifies a monumental stride in what descriptive geometry aims to accomplish. Other AI models working with different spatial representations continue to push the boundaries, offering groundbreaking progress in projective techniques relevant to architectural practice.

Moreover, the application of AI in architecture extends beyond traditional geometric projections. Current models are capable of transforming diverse inputs like sound, video, and text into completely new formats. Examples include converting text into images with AI models like DALL-E, Midjourney, or Stable Diffusion, transforming text into music with AudioLM, and generating video clips from static images with Make-A-Video. Innovative projects further expand these capabilities, like Anna Ridler's Drawing with Sound, Hannah Davis's Symphonologie, and Ross Goodwin's world.camera, showcasing the extensive range of projections possible with today's AI, bridging various domains through creative translations [17].

# 4 Stages of architectural design process

To comprehend the impact of AI at various stages of the architectural design process, an analysis of the classic architectural workflow, from preliminary concept to

**Table 1:** The architectural process from predesign to post-occupancy evaluation

| | |
|---|---|
| *Pre-design/programming:* | Client consultation and establishment of project goals, scope, and budget |
| | Professionals involved: Client, Architect, Interior Designer, and other consultants as needed |
| *Site analysis:* | Evaluation of the site's opportunities, constraints, and context |
| | Professionals involved: Architect, Landscape Architect, Civil Engineer, Environmental Consultant, and other specialists as needed |
| *Schematic design (SD):* | Creation of initial design concepts, sketches, and preliminary space planning |
| | Professionals involved: Architect, Interior Designer, Landscape Architect, Structural Engineer, Mechanical, Electrical, and Plumbing (MEP) Engineers, and other consultants as needed |
| *Design development (DD):* | Further development of the design, refining spatial relationships, materials, and finishes |
| | Professionals involved: Architect, Interior Designer, Landscape Architect, Structural Engineer, MEP Engineers, and other consultants as needed |
| *Construction documents (CD):* | Preparation of detailed drawings and specifications, which will be used for construction bidding, permitting, and construction |
| | Professionals involved: Architect, Interior Designer, Landscape Architect, Structural Engineer, MEP Engineers, and other consultants as needed |
| *Permitting:* | Submission of CDs to local authorities for review and approval |
| | Professionals involved: Architect, General Contractor, and possibly a Permit Expeditor |
| *Construction administration (CA):* | Oversight of construction, ensuring work is completed according to the CDs |
| | Professionals involved: Architect, Client, General Contractor, Subcontractors, and possibly a Construction Manager or Owner's Representative |
| *Construction:* | Execution of the actual building process, including site preparation, foundation work, building erection, and installation of systems |
| | Professionals involved: General Contractors, Subcontractors, and Trade Workers |
| *Inspection and closeout:* | Final inspections, punch list, and completion of any remaining work |
| | Professionals involved: Architect, Client, General Contractor, Building Inspectors, and possibly a Construction Manager or Owner's Representative |
| *Post-occupancy evaluation (POE):* | Assessment of the building's performance, user satisfaction, and identification of any issues to be addressed |
| | Professionals involved: Architect, Client, Building Occupants, and possibly a Post-Occupancy Evaluation Consultant |

Source: Author (2023).

post-construction, was undertaken. Table 1 delineates these stages into key phases, highlighting the involvement of various professionals and stakeholders throughout the process. This analysis draws upon multiple sources that have examined the architectural design process in detail, providing insights into its general workings [18–23].

The architectural process, encompassing an array of stages from initial pre-design and programming to a concluding post-occupancy evaluation, is fundamentally a sequence of labor-intensive tasks, imbued with human expertise. In the preliminary stage of pre-design and programming, architects traditionally collate project specifications and outline the client's aspirations and requirements.

The subsequent stage, site analysis, necessitates a comprehensive exploration of geographical, environmental, and contextual elements influencing the building design. The time-honored method demands on-site visits for data collection and measurement.

The SD phase, a stage heavily reliant on the architect's creativity, intuition, and experience, sees architects generate initial design concepts *via* hand-drawn sketches and physical models. Following this, during the DD phase, architects refine their initial concepts and deliberate over materials and construction techniques. This stage, traditionally involving manual drafting, material research, and consultations with specialists.

The CD phase involves the preparation of coordinated drawings and specifications. Historically, this process has been a manual, iterative one involving the exchange of documents and meetings. The permitting process involves ensuring building code compliance and obtaining relevant permits. Traditionally a manual process, it necessitates the submission of CDs to local authorities and adherence to complex regulations.

In the CA phase, architects coordinate with contractors to ensure work aligns with design intent and CDs. Finally, the post-occupancy evaluation involves assessing building performance and user satisfaction, typically through manual methods such as surveys, interviews, and site visits.

The fundamental workflow from conceptualization to the realization of built structures has remained consistent over the years. Despite this continuity, each phase of the architectural and engineering process has been significantly accelerated and refined through the introduction of new technologies. For instance, the preparation of CDs, historically a time-intensive task requiring manual drafting on drawing tables for extended hours, has been transformed by the adoption of CAD. CAD technology has streamlined this phase, enabling quicker completion without altering the core process itself.

Further advancements were seen with the integration of BIM technologies. BIM has enhanced not only the efficiency but also the accuracy of various stages, including SD, DD, and the creation of CDs. This is largely due to BIM's ability to facilitate a collaborative environment where professionals across disciplines – such as architecture, civil engineering, and electrical engineering – can work concurrently on a unified platform [24]. This collaborative approach significantly reduces errors by ensuring all stakeholders have access to, and work from, a single, shared model. Thus, while the overarching process from design to construction remains unchanged, the introduction of CAD and BIM technologies has fundamentally improved both the speed and precision of these phases.

Nevertheless, now, with the introduction of AI technologies into the field of architecture signifies a pivotal shift toward heightened efficiency, precision, and innovation. This evolution mirrors the transformative impacts witnessed with earlier technological integrations, such as CAD and BIM, which significantly expedited and refined the architectural process without altering its fundamental stages. AI's ability to analyze extensive datasets and generate solutions rapidly surpasses the enhancements brought about by CAD and BIM, offering a profound acceleration of the architectural workflow. For instance, AI applications can automate the intensive drafting tasks of the initial programming stages, and DD tasks that traditionally require substantial time. Moreover, AI's potential in architectural design extends to an array of possibilities through the generative adversarial network (GANs) – a class of machine learning frameworks and a prominent framework for approaching generative AI. These algorithms can explore a vast array of design variations based on predefined criteria, thereby assisting architects in identifying solutions that may not have been immediately apparent through conventional methods. This aspect of AI in architecture is highlighted in the work of researchers such as Leach, who discusses the role of AI in enhancing creative problem-solving within the design process [7].

# 5 AI applications in the architectural process

As outlined previously, throughout history, technology has made significant strides in various stages of the architectural process, from pre-design and programming to CDs. However, with the integration of AI tools many of these architectural processes are already affected. Table 2 is a

**Table 2:** Illustrates the areas where AI has already infiltrated the architectural design process, as previously discussed

| Stage | AI tool/example | Application in architecture | Process |
| --- | --- | --- | --- |
| Pre-Design/ Programming | Varys<br>Midjourney<br>DALL-E 2<br>Chatmind<br>Adobe Firefly<br>Maket<br>Architechtures<br>Designs.ai<br>COMFY UI<br>[27,33–38] | Analyze and optimize space relationships and usage | Manual gathering of project requirements and programming |
| Analysis | XKool<br>Testfit<br>Autodesk Forma<br>Digital Blue Foam<br>Sidewalk Labs<br>Civils.ai<br>Archistar.ai<br>[26,39–44] | Process and analyze geographical and environmental data | Site analysis, including studying geographical, environmental, and contextual factors |
| SD | ReRoom.ai<br>Cove.tool<br>SmartDrew<br>HomeByMe<br>Hutch<br>Programa<br>IKEA Place<br>LexSet<br>ReRender.ai<br>Finch3d<br>Visualise.ai<br>Autodesk Generative Design<br>Interiorai<br>ArchitecHtures<br>RoomGPT<br>Photoshop (Beta)<br>ARK<br>Sloyd<br>ArkoAI<br>RoomGPT<br>ToTree<br>Collovgpt<br>Hestyaai<br>ReimagineHome<br>Homedesignai<br>PromeAI<br>Homestyler<br>Spoak<br>Spacely<br>Fulhaus Ludwig<br>Veras<br>Fabrie<br>Mnml.ai<br>Lookx<br>Getfoorplan<br>Runway.ai<br>[45–78] | Concept generation and design exploration using GANs<br><br>Create images and renders from text descriptions or sketch models<br><br>Generating plans | Sketches, physical models, and design exploration |

**Table 2:** Continued

| Stage | AI tool/example | Application in architecture | Process |
| --- | --- | --- | --- |
| DD | Autodesk's Dreamcatcher<br>Chat GPT<br>Finch3d<br>LEO<br>Hypar.ai<br>nTopology<br>[78–81] | Automate design exploration and optimization processes | DD, material selection, and detailing |
| CDs | Blueprints AI<br>Kreo<br>BricsCAD<br>[82] | BIM automation: design, cost estimation, and construction planning | Preparation of CDs, including drawings and specifications |

Source: Author, 2023.
The selection of applications is based on current usage data.

review until February 2024 on the stages where AI can already be integrated. However, as I am writing this paper, and you are reading it, technologies are being developed at a very fast pace, nonetheless, here is a trial to encompass the changes up until now. The selection of these tools was based on their significance in the industry, creative capabilities, and representation in the current architectural industry, with a total of over 70 AI tools were examined during the study.

The representation phase – which is from preliminary design to the DD – had seen the greatest advances with AI, and this is due to machine advances in image generation, it is now feasible to generate visually impressive and hyperreal visuals, which can assist designers in envisioning their design concepts in a new way. Two options are easily accessible to everyone: text-to-image or image-to-image. In the text-to-image approach, users have the option to either select a predefined selection of labels or provide a comprehensive description/prompt. Using the provided input, the AI then produces diverse visuals that function as design thoughts or inspiration sources. The swift transformation of keywords or guidelines into visual representations greatly aids designers in the examination and experimentation of their concepts. The process of image-to-image generation allows users to input a wide variety of image formats, including screenshots of 3D massing, hand-drawn sketches, realistic photographs, and architectural plan drawings. These images can then be used as reference points for composition. In addition, users have the option to select an extra image to be used as a source of inspiration or as a reference for the desired "processing." This approach is additionally enhanced by prompts specifically designed to improve the machine's understanding of project requirements, thereby making it easier to generate images that closely match the input [25].

This section provides an overview of each phase within the architectural process, highlighting the role of AI and showcasing a specific tool that exemplifies AI's transformative impact on the industry, whether it is text-to-text, image-to-text, or something else. We will look into how AI is integrated at various stages, throughout the architectural process. By examining a particular AI tool in action, we aim to illustrate the substantial shifts AI is already bringing to the architectural workflow. However, many of the tools mentioned do not only fit in one of the categories, actually they might be used at several stages and not only one.

To start with, during the pre-design and programming phase, *Maket* utilizes advanced AI algorithms to understand project constraints, allowing users to specify land sizing, building dimensions, and desired room adjacencies. This flexibility enables the generation of floor plans that adhere to project requirements. Its generative AI capabilities empower users to test concepts swiftly, producing thousands of early-stage design options within minutes. This feature is particularly valuable for exploring various architectural solutions and selecting the best fit for the project's vision. Furthermore, Maket enhances the visualization of designs by allowing users to quickly transition from 2D floor plans to 3D models (which is also helpful during the DD phase) (Figure 3).

During the analysis stage of architectural design, tools like *Digital Blue Foam* facilitate sun and wind studies, allowing designers to generate and download reports containing critical information for the project [26]. This tool enables quick activation of sun and wind analysis options directly from its interface. Specifically, it offers functionalities for shadow analysis, sun hours calculation, and daylight autonomy evaluation, which are crucial for assessing a project's environmental impact and sustainability. Wind studies within Digital Blue Foam visualize wind direction and prevalence, providing essential data for optimizing building orientation and design for natural ventilation. Furthermore, the platform is designed for generating sustainable building designs by incorporating real-time Daylight Autonomy Score, Solar Radiation Simulation, Instant Shadow Study, Wind Score, and Sun Path Animation. These features allow designers to validate designs in real-time, aiming for net-zero $CO_2$ buildings without relying on expensive, complex software. The tool also introduces a proprietary neighborhood scoring system that supports the "15-minute city" planning strategy, emphasizing sustainability and accessibility to amenities (Figure 4).

In the SD and DD phases, there has been a notable increase in the development of applications aimed at enhancing these stages. While Table 2 attempts a comprehensive review, it is evident that numerous other applications could significantly contribute to these categories.

*ArchitecHtures*, for example, accelerates the design process by allowing architects to input design criteria and rapidly receive optimized solutions. This tool streamlines feasibility analysis, encourages rapid iteration, and provides accurate insights, ultimately making the transition to BIM models more efficient (Figure 5). Many of these tools at this stage represent a significant advancement, as they offer speed, precision, and creativity in those design phases [27].

The CDs phase represents a critical juncture where the conceptual and SDs transition into detailed, actionable blueprints that guide the actual building process. This stage is pivotal for translating design intent into technical specifications that include detailed drawings, material specifications, and documentation necessary for construction bidding, permitting, and building. A good example of an AI tool that can now aid architects is Swapp, which is an AI-powered tool designed to create CDs. Swapp uses advanced algorithms to analyze previous projects, extracting and learning from the design habits and annotation practices embedded within past documentations. This analysis enables Swapp to develop a customized set of rules and algorithms that automates the generation and annotation of architectural documentation for new projects [28].

At the core of Swapp's functionality is its design decision language, a system that encapsulates the foundational rules of architecture and applies them to the automated creation of CDs. By analyzing previous Revit models, Swapp formulates "recipes" or templates that can be applied to new projects. This AI-driven process automates the generation of detailed, code-compliant Revit models and also set styles, and creates a complete set of construction drawings.

In the Project Management/CA phase, ALICE Technologies, for example, is also an AI tool that aids in easing the process in that phase. This tool is designed for owners of large-scale commercial, industrial, or infrastructure projects, ALICE optimizes construction planning, sequencing, and scheduling. This tool allows for efficient assessment, rapid prototyping of ideas, and enhanced forecasting, enabling better financial projections and the mitigation of risks. By exploring numerous potential ways to build a project before breaking ground, ALICE empowers stakeholders to make informed decisions, optimizing resource management and reducing project risks and costs [29].

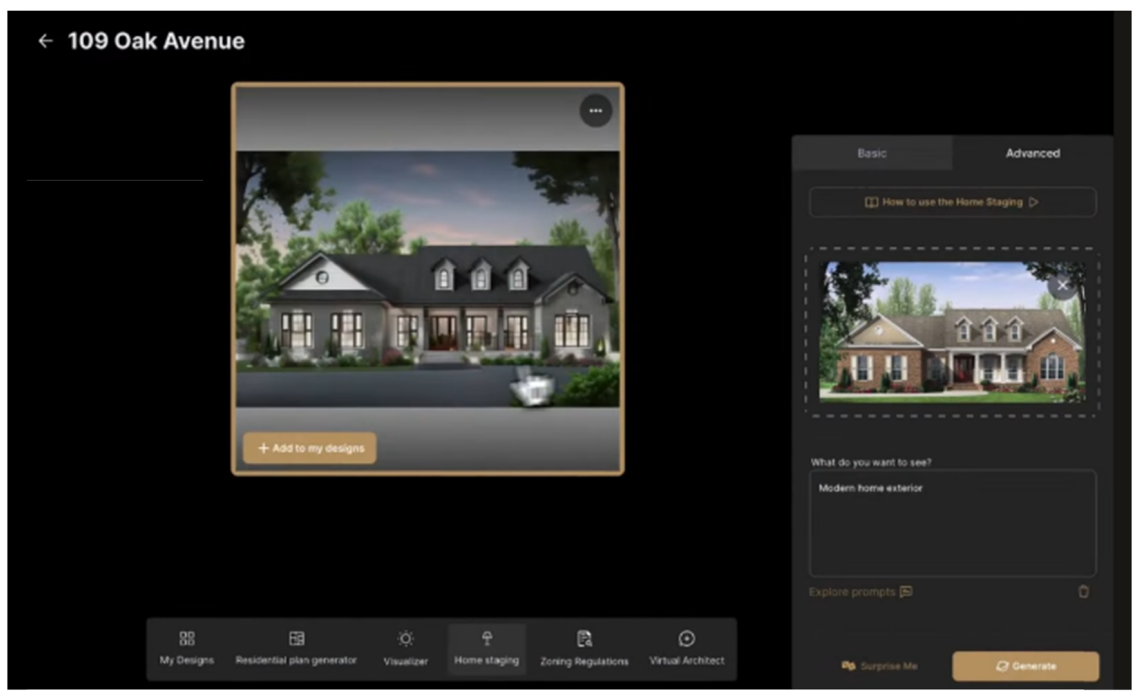

**Figure 3:** Maket, from prompt to various designs. Source: Maket Website, February 14, 2024 [37].

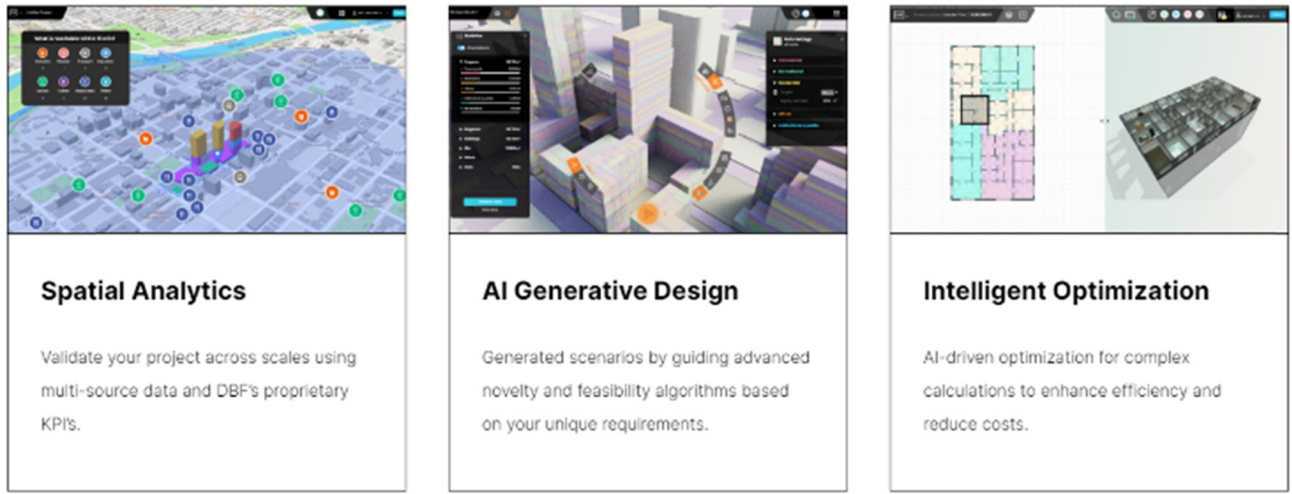

**Figure 4:** Digital Blue Foam platform on several aspects they can perform in analysis. Source: Digital Blue Foam Website, February 14, 2024.

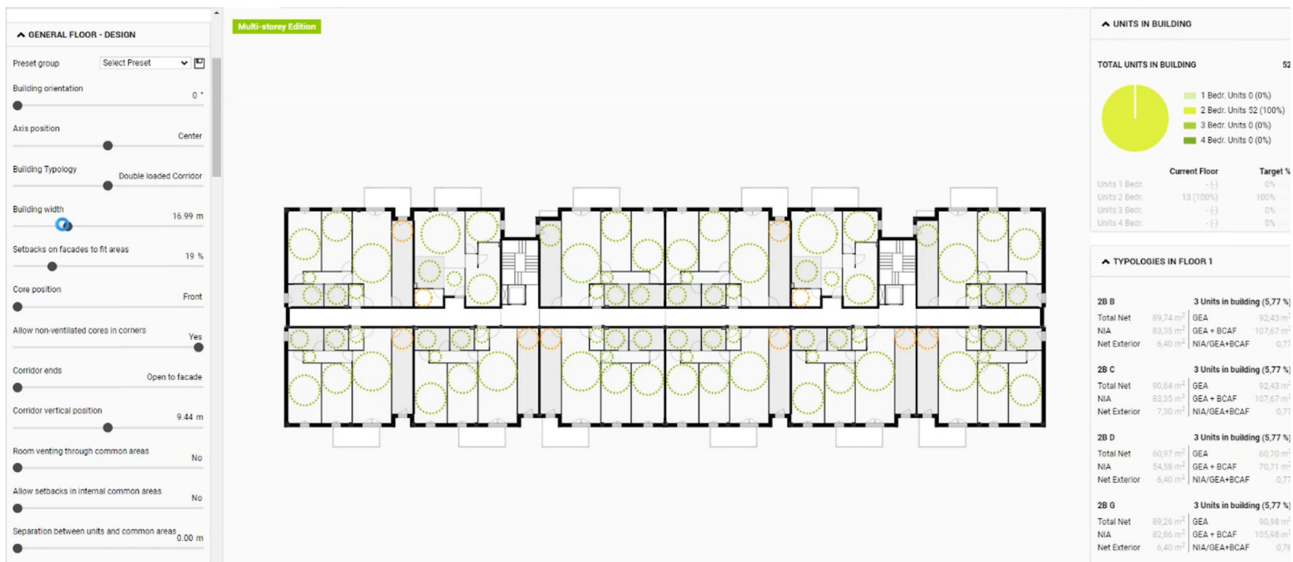

**Figure 5:** ArchitecHtures in the DD phase. Source: ArchitecHtures Website, February 14, 2024.

Finally, throughout the construction lifecycle, including post-occupancy evaluation, AI and robotics have started to play a role, albeit in early development stages. Tools like Canvas, which precisely applies interior plaster, and Construction Robotics' Mule, which aids in lifting, show potential. However, their effectiveness is currently limited by the need for significant human intervention for setup and operation. Similarly, 3D house printing technologies are emerging, indicating a trend toward automation, but still require development to reduce labor dependency [30–32].

## 6 Magnitude of change in the architecture process – discussion

As demonstrated in Table 2, AI tools are already integrated into every stage of the architectural design process. Notably, during the representation phase – spanning from preliminary design to the DD phase – there is a significant increase in the creation of tools. This surge can be attributed to the widespread availability of diffusion models. In this process, diffusion models can alter architectural materials, styles, elements, and forms. How does this assist in design scenarios? After case studies and design directions or preliminary sketches have been determined, it becomes feasible to quickly convert design keywords or sketches into visual depictions of spatial ideas. Moreover, it allows for the quick creation of hyper realistic visual pictures that offer a more intuitive expression of design ideas, without requiring intricate geometric models. This technique has the ability to enhance the variety of possibilities by making different modifications to a single chosen image [83].

Additionally, if we look into the history of text-to-image models, the precision of these visual-based algorithms has evolved significantly, capturing more complex and structured ideas. Advanced models, including diffusers and LLMs, encode sophisticated abstractions beyond the basic elements seen in parametric modeling. The latest text-to-image models illustrate this advancement, converting intricate text prompts into high-quality visuals by referencing styles, historical periods, and artistic movements. Moreover, the rate of advancement in these research domains is remarkable. When we input the same instructions into these models, it is evident how quickly these technologies have started to mimic and enhance concepts that are fundamental to our field. A comparison, as shown in Figure 6, from the GLIDE model (December 2021) to Midjourney V6 (February 2024), reveals a significant enhancement in the realism of outputs, like floor plans, within a mere year and a half [17].

To illustrate the magnitude of change in the representation stage, Damilee[1] conducted a simple experiment, where three human architects were juxtaposed against AI tools in a contest to design a 280-square-meter abode, situated on the challenging contours of a sloped terrain. As the designs unfolded, each encapsulating distinct perspectives and subtleties, the Archibeans Discord community was beckoned to judge, blind to the originator of each design. The results of the designs conducted by an AI tool won most of the votes; nonetheless, the velocity with which these designs were generated outpaced human counterparts, 62% of the voters chose the anonymized design created by AI, while 38% chose the anonymized design created by human trained architects.

The AI's prowess was not confined to mere design; it also showed in the rendering and textual challenges, underscoring its growing prominence in architectural practices.

Following the experiment by DamiLee, the implications of such revelations are far-reaching. This shows the paradigm shift in the architectural realm we are witnessing. Nevertheless, it is clear that the spatial quality of the outcomes by human architects is yet significantly inferior to those generated by AI. This was particularly noticeable when comparing the plans created by an AI tool with the spatial quality and depth of thought evident in the works produced by architects. This experiment was conducted again in late 2024, and AI won by 81% of the community votes.[2]

Additionally, AI systems like Google's DeepMind have also demonstrated their utility in optimizing building performance, notably achieving a 40% reduction in energy consumption within data centers [84]. This represents more than a technical achievement – it reflects a changing role for architects. AI is no longer a simple tool at their disposal but a collaborator, offering data-driven insights that reshape how buildings are conceptualized and realized.

However, yet, a recent study on public perceptions of AI's design abilities reveals a more complex picture. Despite the clear technological advantages AI offers, many still believe that AI underperforms in areas that require nuanced, user-specific considerations. Interestingly, the more knowledgeable individuals perceive themselves to be about design, and the older they are, the more likely they are to believe that AI might surpass human designers in some respects. This ambivalence highlights the evolving nature of AI acceptance in design fields. While AI's speed and efficiency are undeniable, the perception of its ability to address human needs remains a point of tension, suggesting that the integration of AI into architectural practice may still require a broader cultural shift [84].

Phil Bernstein, in his article, identifies three distinct possibilities for the coming era of machine autonomy in architecture. He envisions AI rationalizing the vast amounts of digital data emerging from the current interstice, once known as "interoperability." Second, an obsession with image creation, form-finding, and unrestrained exploration of aesthetic options. Lastly, he sees AI being utilized for the generation and analysis of technical and performative aspects of building design, making results more valid, explainable, and responsible [85].

---

[1] We tried to compete with AI... [AI vs ARCHITECT] – DamiLee https://www.youtube.com/watch?v=N709ZrxoIP0&t=16s&ab_channel=DamiLee.

[2] https://www.youtube.com/watch?v=bvhHzaq5iKk&list=WL&index=1&ab_channel=DamiLee.

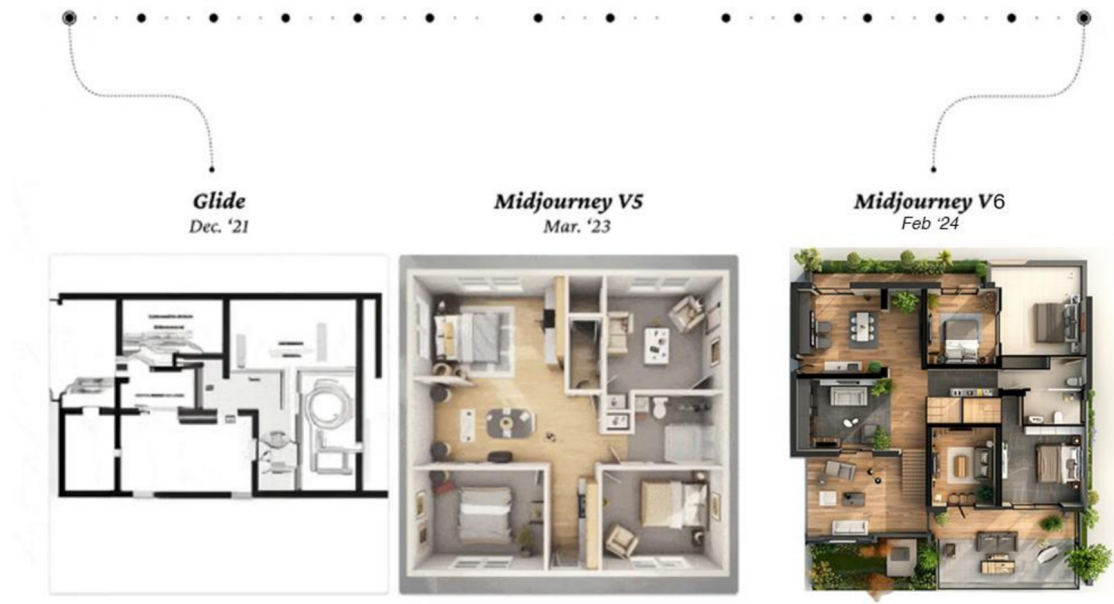

**Figure 6:** The difference in 18 months of evolution in creating a plan by AI. Source: Stanislas and Author (2024).

Nonetheless, as AI tools, such as image generators using GAN technologies and diffusions, inspire new form explorations, machine learning analysis can evaluate, prioritize, and generate further options substantiated for clients. Tools like Digital Blue Foam exemplify this capability, and with an increasing number of clients accessing these tools online – either for free or through paid services – to create their own designs without prior experience (or relying on a trained architect/designer), a fundamental question arises: If AI can – although not fully – handle aesthetics, data organization, and design optimization, what roles and capabilities remain for future human architects? As we are still just discovering these tools. This question challenges the profession to redefine its role in a landscape increasingly shaped by advanced technologies.

However, despite these advancements mentioned, the essential aspects of architectural design that necessitate an architect's unique insight remain predominantly untouched by AI. These aspects encompass spatial comprehension and the integration of diverse scientific knowledge to address societal needs. Critical thinking plays a pivotal role in decision-making and problem identification. As demonstrated in Figure 7, AI tools have so far primarily dominated the representation phase. However, the stages where data is collected, processed, and critically analyzed – drawing upon extensive background knowledge to arrive at solutions – remain largely within the domain of the human architect. While AI is indispensable at this stage for processing data such as material usage, sun orientation, and wind patterns, the architect must ultimately apply critical thinking to synthesize this information and devise solutions. Once a solution is reached, AI's role in the representation phase is far more advanced and faster than any human capability. This reemphasizes the architect's role underscoring the intrinsic value of their contribution to the field. Architects of the future are called upon not merely as creators of visual artifacts but as curators of experiences deeply embedded in the philosophical, historical, social, and political contexts of their projects. Their expertise becomes crucial in leveraging AI tools not for the sake of automation alone but as instruments in a larger process of problem-solving, meaningful creation, "data optimization," and enhancing sustainable outcomes. This perspective elevates the importance of focusing on the existential aspects of dwelling, transcending the limitations of an ocular-centric approach to architecture. It emphasizes the need for architects to engage with the existential needs of living spaces, thereby ensuring that technological advancements like AI serve to enhance, rather than diminish, the human-centric goals of architecture. Therefore, architects can evolve from passive users of technological tools into active beings of information. They can be proficient in navigating multiple domains while adeptly managing the available tools to develop innovative solutions that address societal needs. This perspective positions architects at the forefront of future job demands, emphasizing their central role in shaping a better future rather than marginalizing their abilities.

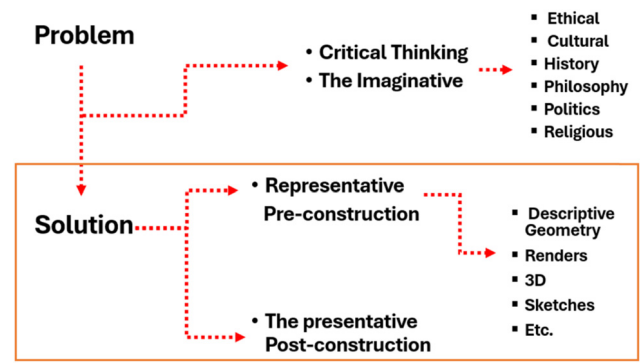

**Figure 7:** The problem-solution phases in architecture. Source: Author (2024).

However, ethical questions still arise in terms of labor displacement and authorship. As AI tools become more adept at performing tasks traditionally done by humans, it raises deeper questions about the future of the profession and the value of human creativity. If AI can produce better, faster, and cheaper designs, what place is left for the architect as an artisan and thinker? Moreover, the reliance on AI to gather and analyze data introduces risks of bias and exclusion, particularly when AI systems draw on unregulated or biased datasets (or pre-trained models). Additionally, if AI is doing most of the work, will architects still hold on to the "architects-ego" of sole creators, or will the work be labeled more of a collaborative entity for the further betterment of society?

# 7 Conclusion

This article presents an exploration of the integration of AI into the architectural design process, emphasizing the significant transformative potential AI offers while acknowledging the challenges it introduces. The review underscores the relationship between human creativity and AI's computational capabilities, highlighting the need for a collaborative approach to utilize AI's benefits effectively. As we have seen, AI tools are already now integral to every stage of the architectural design process, particularly during the representation phase, where they enable quick conversion of sketches into realistic visual depictions, among other capabilities. This has led to a significant enhancement in design possibilities and efficiency. Additionally, text-to-image models have evolved rapidly, producing high-quality visuals that mimic and enhance fundamental architectural concepts. These advancements highlight a paradigm shift in architecture, where AI's capabilities in design, rendering, and analysis are surpassing traditional methods. The implications are profound, suggesting AI's potential to make architectural design more efficient, innovative, and data-driven, while challenging architects to redefine their roles in this evolving landscape.

Moreover, the discussion extends to the dominance of visual aesthetics in architectural practice, a trend deeply rooted in historical precedence but now challenged. The ocular-centric approach, which prioritized visual impact, finds a modern counterpart in the way architecture is often consumed and appreciated today. However, the advent of AI in architectural design prompts a critical reassessment of this tendency. While AI excels in generating visually compelling designs, especially in the representative phase, as AI is capable of completing these redundant tasks at unprecedented speeds and accuracy, however, its current limitations in fully grasping the multisensory and experiential qualities of space. This era of AI integration, therefore, calls for a balanced approach that values not only the visual but also the tactile, the auditory, and the spatial analysis that well trained experienced architects can offer, the part that is focused on the critical thinking part of architecture to find the solution. Therefore, architects must leverage this technology, not as a mere tool for efficiency but as a collaborative partner in exploring new design paradigms, looking into the existential, and much-needed precise sustainable practices for a better society. This approach entails a shift towards creating spaces that resonate on a deeper, more human level, moving beyond the purely retinal to encompass a holistic experience. A shift in focus is required for designers, emphasizing a deeper integration of humanities education – encompassing philosophical, historical, political, economic, sociological perspectives, *etc.* – into the design process. By using AI tools to streamline and automate routine tasks, designers can free up time to concentrate on these critical aspects, the existential. This approach positions designers as curators of objects and spaces, rather than solely as creators, thereby enriching their work with a broader and more profound contextual understanding, that is going to be much needed in the future.

**Funding information:** This work was supported by the Open Access Publishing Fund of the Free University of Bozen-Bolzano.

**Author contribution:** The author confirms the sole responsibility for the conception of the study, presented results, and manuscript preparation.

**Conflict of interest:** The author states no conflict of interest.

**Data availability statement:** Most datasets generated and analyzed in this study are comprised in this submitted manuscript. The other datasets are available on reasonable request from the corresponding author with the attached information.